\documentclass[%
 reprint,
nofootinbib,
longbibliography,
 amsmath,amssymb,
 aps,
prl,
twocolumn,
superscriptaddress
]{revtex4-2}

\usepackage{lineno} 
\usepackage[pdftex]{graphicx}
\usepackage{subfigure} 
\usepackage{epstopdf}
\usepackage{graphicx}
\usepackage{dcolumn}
\usepackage{bm}
\usepackage[colorlinks,
linkcolor=blue, citecolor=blue, anchorcolor=blue,
urlcolor=blue,
]{hyperref}
\usepackage{siunitx}
\usepackage{braket}
\usepackage{color}
\usepackage{tikz}
\usepackage{textcomp}

\newcommand{\Yb}{$^{171}\textrm{Yb}^+~$}
\newcommand{\Ba}{$^{138}\textrm{Ba}^+~$}

\newcommand{\affa}{State Key Laboratory of Low Dimensional Quantum Physics, Department of Physics, Tsinghua University, Beijing 100084, China}
\newcommand{\affHefei}{Hefei National Laboratory, Hefei 230088, China}
\newcommand{\affRUC}{School of Physics and Beijing Key Laboratory of Opto-electronic Functional Materials and Micro-nano Devices, Renmin University of China, 100872 Beijing, China}
\newcommand{\affBAQ}{Beijing Academy of Quantum Information Sciences, Beijing 100193, China}
\newcommand{\affFront}{Frontier Science Center for Quantum Information, Beijing 100084, China}
\newcommand{\affIBS}{Center for Trapped Ion Quantum Science, Institute for Basic Science, Daejeon 34126, South Korea}

\begin{document}

\title{Beyond-Ten-Hour Coherence in a Decoherence-Free Trapped-Ion Clock Qubit}

\author{Jiahao Pi}
\thanks{These authors contributed equally to this work.} 
\affiliation{\affa}

\author{Xiangjia Liu}
\thanks{These authors contributed equally to this work.} 
\affiliation{\affa}
\author{Junle Cao}
\affiliation{\affa}

\author{Pengfei Wang}
\affiliation{\affBAQ}
\affiliation{\affHefei}

\author{Lingfeng Ou}
\author{Erfu Gao}
\author{Hengchao Tu}
\author{Menglin Zou}
\affiliation{\affa}

\author{Xiang Zhang}
\affiliation{\affBAQ}
\affiliation{\affRUC}

\author{Junhua Zhang}
\affiliation{International Quantum Academy, Shenzhen, 518048, China.}
\author{Kihwan Kim}
\affiliation{\affa}
\affiliation{\affBAQ}
\affiliation{\affHefei}
\affiliation{\affFront}
\affiliation{\affIBS}


\begin{abstract}
Quantum systems promise to revolutionize information processing science and technology \cite{Wiesner83,bennett1984quantum,shor1994algorithms}. The preservation of quantum coherence, the defining property of qubits, fundamentally constrains the performance of quantum information processing with quantum memories \cite{divincenzo2000physical}. While trapped atomic ions theoretically support million-year coherence based on spontaneous emission \cite{wild1952radio,4065250,pritchard201221}, experimental demonstrations have reached far less, only about an hour~\cite{bollinger91a,fisk95very,Langer2005,harty2014high,YeWang17,wang2021single}. Here we combine clock-state qubits with decoherence-free subspace (DFS) encoding to achieve coherence exceeding ten hours. Using correlation-based phase tracking in \Yb ion pairs sympathetically cooled by \Ba ion, we demonstrate this without magnetic shielding or enhanced microwave phase stabilization that previously limited coherence times. DFS encoding references the qubit phase to the inter-ion energy difference to reject microwave phase noise and common-mode magnetic fluctuations, while clock states provide environmental insensitivity. Throughout  measurements extended to 1600 seconds, we observe minimal coherence decay, with exponential fits yielding a coherence time of $(3.77 \pm 1.09)\times 10^{4}$ seconds. Our results establish DFS encoding as a form of passive error correction that eliminates technical noise constraints, unlocking the million-year coherence potential of atomic ions for scalable quantum information processing.

\end{abstract}

\maketitle


As quantum technologies mature, it is natural to ask what constitutes the ultimate qubit~\cite{kim2014trapped}, the fundamental building block for quantum information processing~\cite{divincenzo2000physical,Wiesner83,bennett1984quantum,shor1994algorithms}. It should, above all, allow quantum information to be written and read out reliably, and the ability to preserve quantum information, particularly quantum coherence, as long as possible, is the most fundamental and defining attribute~\cite{divincenzo2000physical}. From a fundamental perspective, the hyperfine ground states of single atoms represent ideal qubit candidates, particularly clock states that exhibit minimal sensitivity to environmental perturbations~\cite{bollinger91a,fisk95very,Langer2005,harty2014high,YeWang17,wang2021single}. Indeed, trapped \Yb ions have shown the longest coherence times to date, reaching 90 minutes~\cite{YeWang17,wang2021single}. Yet this falls dramatically short of the theoretical limit, considering only spontaneous emission, coherence should persist for millions of years~\cite{wild1952radio,4065250,pritchard201221}.

In trapped-ion systems with effectively unbounded storage times, hyperfine ground-state clock qubits have routinely shown coherence times of seconds to one minute in multiple systems~\cite{Langer2005,harty2014high}. This limit arises from motional heating during interrogation, which degrades state discrimination~\cite{Epstein07,wesenberg2007fluorescence,kotler2014measurement}. Sympathetic cooling with a different species of co-trapped atomic ions extends coherence to 10 minutes by maintaining detection efficiency~\cite{YeWang17}. Even here, the dominant bottleneck was not intrinsic qubit decoherence, but rather the phase stability of the interrogating microwave oscillator, where improved oscillators enabled coherence times of up to 90 minutes~\cite{wang2021single}. Nevertheless, extending coherence further requires overcoming both ambient magnetic field fluctuations and the residual phase noise of the microwave oscillator. To push beyond this regime, one approaches the domain of quantum error correction (QEC)~\cite{ni_beating_2023,sivak_real-time_2023}. While active QEC requires significant resource overhead, a powerful passive implementation is to encode the logical qubit in a decoherence-free subspace (DFS). By defining the qubit via the energy difference between closely spaced ions, this approach yields a vanishing energy splitting that is inherently immune to uniform magnetic field fluctuations and the global phase noise of the microwave oscillator~\cite{Palma1996, Lidar1998, Zanardi1997, Kielpinski2001}.   

DFS encoding has previously extended Zeeman-qubit coherence from milliseconds to tens of seconds~\cite{Kielpinski2001,Roos2004,Haffner05} and enabled high-precision quantum metrology~\cite{roos2006designer,chwalla_precision_2007,kotler2014measurement,pruttivarasin2015michelson}. Here, we combine clock qubits with DFS encoding and employ correlation-based phase-tracking methods~\cite{chwalla_precision_2007,pruttivarasin2015michelson} to demonstrate coherence exceeding ten hours without implementing magnetic field shielding and improving microwave references~\cite{wang2021single}. The approach exploits the simultaneous cancelation of common-mode magnetic fluctuations over micrometer-scale separations and global microwave phase noise, yielding remarkable technical simplicity and robustness. We realize the DFS logical qubit using two $^{171}\text{Yb}^+$ ions sympathetically cooled by a central $^{138}\text{Ba}^+$ ion. In measurements extending to 1,600 seconds, correlation contrast shows only minimal decay within experimental uncertainty, with fitted coherence times exceeding  $3.77\times 10^{4}$ seconds. Residual decoherence is dominated by dephasing arising from exchange hopping of the \Yb ion positions (see Methods).

\begin{figure*}[htbp]
\centering
\includegraphics[width=0.95\textwidth]{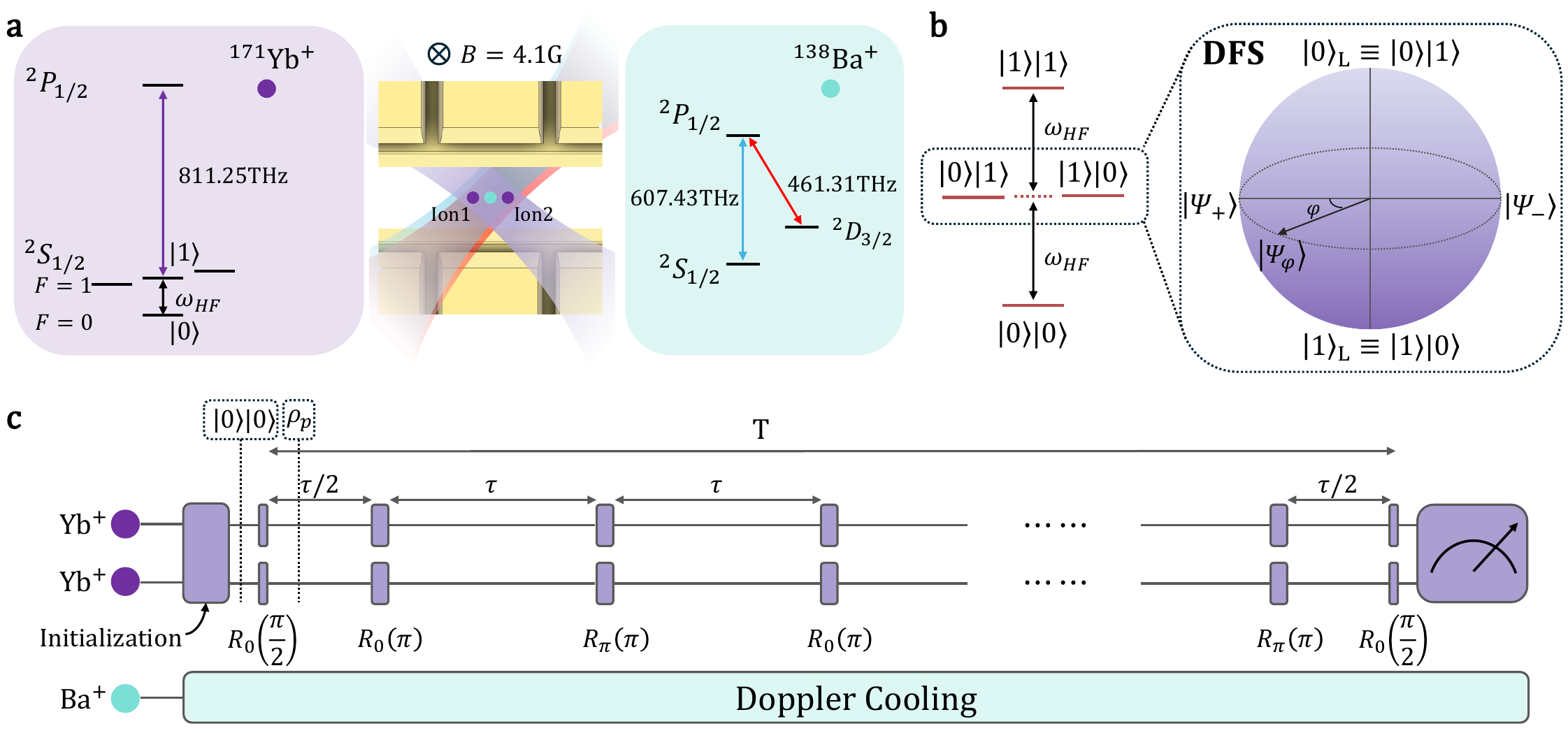} 
\caption{\textbf{Extending Coherence Time of a $^{171}\text{Yb}^{+}$ DFS Qubit via $^{138}\text{Ba}^{+}$ Sympathetic Cooling. }
\textbf{(a)} Energy level diagrams for the $^{171}\text{Yb}^{+}$ qubit ion and the $^{138}\text{Ba}^{+}$ coolant ion. The $^{171}\text{Yb}^{+}$ qubit is encoded in the hyperfine clock transition of the ground state ($F=0 \leftrightarrow F=1$) with a splitting of $12.64\,\text{GHz}$. The $^{138}\text{Ba}^{+}$ ion is used for sympathetic cooling via the visible transitions shown (corresponding to $493\,\text{nm}$ and $650\,\text{nm}$). The central schematic depicts the linear Paul trap confining a mixed-species ion chain (Yb-Ba-Yb) subjected to a quantizing magnetic field of $B = 4.1\,\text{G}$. 
\textbf{(b)} Logical qubit encoding. The logical qubit $\{\ket{\textbf{0}}_{\rm L}, \ket{\textbf{1}}_{\rm L}\}$ is encoded in the decoherence-free subspace (DFS) spanned by the states $\{\ket{0}\ket{1}, \ket{1}\ket{0}\}$, providing robustness against collective magnetic noise. The general  state $\Psi_{\varphi}$ is visualized on a logical Bloch sphere.
\textbf{(c)} Pulse sequence for the coherence time measurement. After initialization and a rotation $R_{0}{\left(\frac{\pi}{2}\right)}$, a dynamical decoupling sequence with total duration of $\rm T$, composed of reverse style spin-echo blocks ($R_0(\pi), R_\pi(\pi)$) with waiting time $\tau$, is applied to the two $^{171}\text{Yb}^+$ qubits to measure the coherence of the logical state. After single ion decoherence, $\rho_{p}$ (Eq. \ref{eq:DFS}) is prepared. Throughout the whole sequence, continuous Doppler cooling is applied to the $^{138}\text{Ba}^{+}$ ion (sympathetic cooling) to suppress thermal motion.}
\label{fig1} 
\end{figure*}

In our experiment, the physical qubits are encoded in the two hyperfine clock states of the $^{171}\text{Yb}^+$ ground $S_{1/2}$ manifold, defined as $\ket{0} \equiv \ket{F=0, m_F=0}$ and $\ket{1} \equiv \ket{F=1, m_F=0}$, separated by \(\omega_{\rm HF}=2\pi(12.642812118 + 310.8\, B^{2})\ \mathrm{GHz}\)~\cite{Olmschenk2007,YeWang17,wang2021single} as shown in Fig.~\ref{fig1}a, where $B$ is the  magnetic field strength. We confine a mixed-species linear ion chain in a symmetric Yb-Ba-Yb configuration within a Paul trap, where ion distance is $6$ \textmu m, subjected to a magnetic field of $B=\SI{4.1}{G}$, shown in Fig.~\ref{fig1}a. We encode a logical qubit within the decoherence-free subspace (DFS) spanned by the anti-aligned states $\{\ket{\textbf{0}}_{\rm L} \equiv \ket{0}\ket{1}, \ket{\textbf{1}}_{\rm L} \equiv \ket{1}\ket{0}\}$, as illustrated in Fig.~\ref{fig1}b.

Our sympathetic cooling scheme (Fig.~\ref{fig1}a), employing a dedicated $^{138}\text{Ba}^+$ coolant ion, is essential for maintaining long-term coherence, as direct laser cooling of the $^{171}\text{Yb}^+$ qubits would destroy their quantum state. The choice of $^{138}\text{Ba}^+$ is optimal because its cooling transitions are far-detuned from any resonant transitions in $^{171}\text{Yb}^+$ , and the similar atomic masses ensure efficient energy transfer via Coulomb interaction~\cite{Inlek2017,YeWang17,wang2021single}.

The experimental sequence is illustrated in Fig.~\ref{fig1}c. Following initialization to the $\ket{0}\ket{0}$ state, a global $R_0(\pi/2)$ microwave pulse prepares the initial superposition. For evolution times exceeding the single-qubit coherence time, microwave noise causes the coherence of individual ions to decay. Nevertheless, the coherence within the decoherence-free subspace (DFS) is preserved, and the system evolves into a mixed state described by the density matrix~\cite{chwalla_precision_2007}:
\begin{equation}
    \rho_p = \frac{1}{2} |\Psi_\varphi\rangle\langle\Psi_\varphi| + \frac{1}{4} |00\rangle\langle 00| + \frac{1}{4} |11\rangle\langle 11|
    \label{eq:DFS}
\end{equation}
where the coherent component, $\ket{\Psi_\varphi} = \frac{1}{\sqrt{2}} (\ket{0}\ket{1} + e^{i\varphi}\ket{1}\ket{0})$, represents the entangled state that has accumulated a relative phase $\varphi$ due to magnetic field inhomogeneities between the ions (Fig.~\ref{fig1}b). To counteract dephasing from phase fluctuation and other low-frequency noise, a dynamical decoupling sequence comprising reverse style spin-echo blocks is applied. Each block ($R_0(\pi), R_\pi(\pi)$) with waiting time $\tau$ reverses phase accumulation ($\varphi$) and refocusses it. At the end of the sequence, a final global $R_0(\pi/2)$ pulse is applied for analysis. The final quantum state of each ion is read out via state-dependent fluorescence~\cite{Olmschenk2007} with readout error mitigation~\cite{shen_correcting_2012}. From the measurement outcomes,  correlation-based phase tracking is implemented by calculating the two-qubit parity, $P = \langle \sigma_{z_{1}} \sigma_{z_{2}} \rangle$. The coherence time is determined by measuring the exponential decay of this parity signal over time. We note that the contrast is limited at 0.5, consistent with the statistical bound of utilizing a mixture where only the DFS manifold $(\ket{0}\ket{1},\ket{1}\ket{0})$ contributes to the coherent signal.

\begin{figure}[htbp]
\centering
\includegraphics[width=0.45\textwidth]{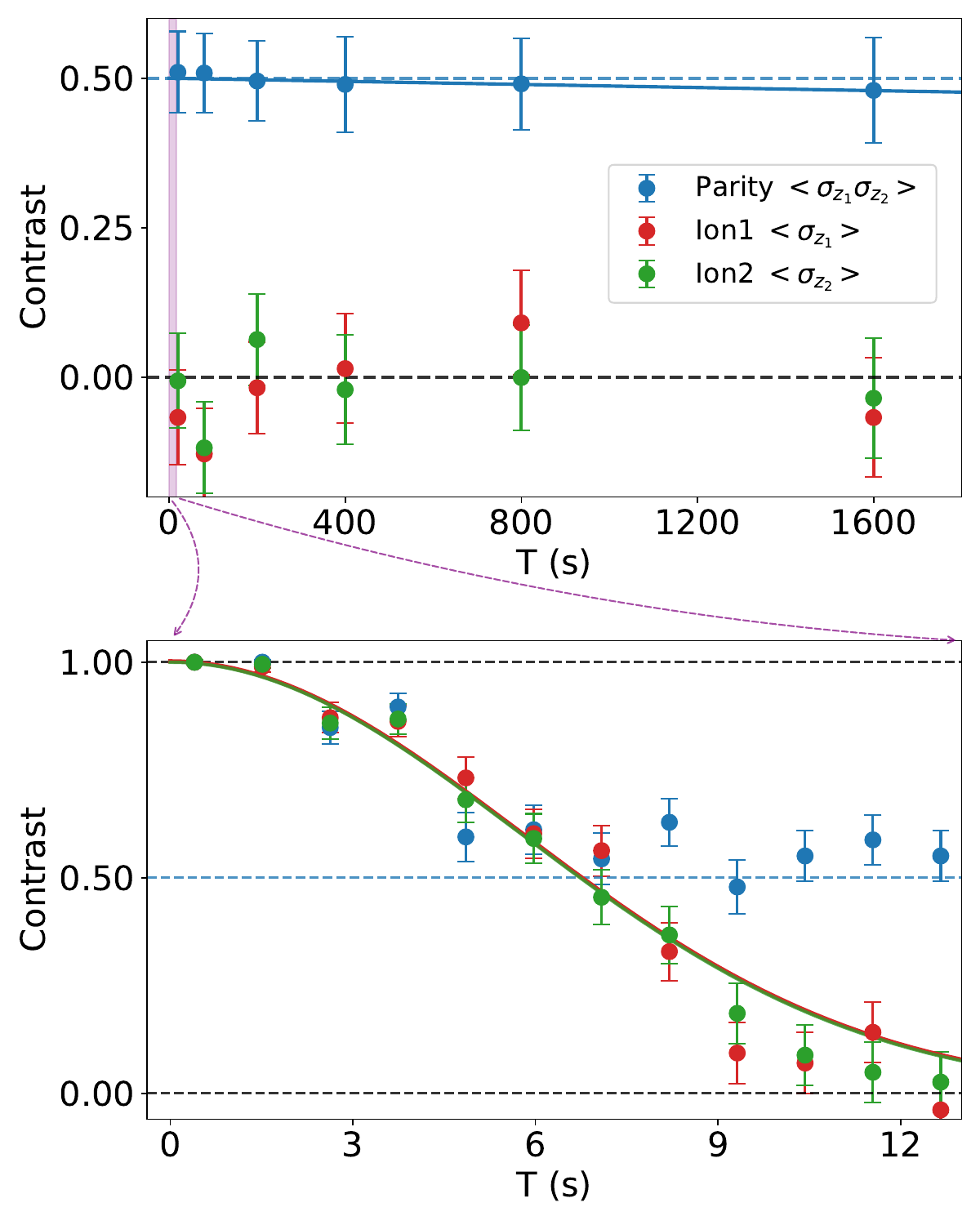}
\caption{\textbf{Coherence time of DFS qubit.} \textbf{(Top)} Long-term evolution of the Ramsey parity contrast extended up to $1600\,\text{s}$. The parity  contrast (blue circles) representing the DFS qubit coherence, remains stable near its maximum value of 0.5(dashed blue line).
 A fit(solid blue line)  to the full dataset yields a coherence time of $\rm (3.77 \pm 1.09)\times 10^{4}~\mathrm{s}$. The Ramsey contrasts of individual ions (red and green circles)vanish effectively to zero, indicating complete decoherence of the physical qubits (mean values of $-0.029 \pm 0.032$ and $-0.019 \pm 0.032$). \textbf{(Bottom)} Zoomed-in view of the first 12s, indicated by the purple shaded region of top figure. The panel shows the rapid decoherence of physical qubits, with coherence times of $\rm T_{Ion1} = 8.10 \pm 0.27$\,s and $\rm T_{Ion2} = 8.09 \pm 0.27$\,s by gaussian decay fit(solid red and green lines). Beyond this decoherence timescale, the state $\rho_p$(Eq. \ref{eq:DFS}) is effectively prepared.} 
\label{fig:clock_coherence}
\end{figure}

The coherence properties of the DFS clock qubit are experimentally characterized, with the results summarized in Fig.~\ref{fig:clock_coherence}. As shown in the top panel, the parity signal exhibits exceptional coherence preservation, decaying by only $\sim 4\%$ over an extended duration of $1600~\text{s}$. An exponential fit to the data yields a coherence time for the logical qubit $\{\ket{\mathbf{0}}_{\mathrm{L}}, \ket{\mathbf{1}}_{\mathrm{L}}\}$ of $\rm T_{\mathrm{DFS}} = (3.77 \pm 1.09)\times 10^{4}~\mathrm{s}$
, where the uncertainty arises from the fitting statistics. This corresponds to approximately $10.5~\text{hours}$, representing an order-of-magnitude improvement over previous results~\cite{wang2021single,YeWang17}. To ensure statistical significance, each data point in this measurement represents the average of 150 to 200 experimental repetitions. The measurements utilize a hybrid dynamical decoupling protocol: a single reverse style spin-echo block is applied for short evolution times ($\rm T < \SI{200}{s}$), while for longer durations ($\rm T \geq \SI{200}{s}$), the sequence is constructed from multiple blocks with a fixed interval of $\tau = \SI{100}{s}$ as shown in Fig.~\ref{fig1}c.

The validity of this long-time coherence is explicitly supported by the dynamics of the individual physical qubits. As seen in the bottom panel of Fig.~\ref{fig:clock_coherence}, under the influence of ambient magnetic field noise, the single-ion coherence decays rapidly with a characteristic time of $\rm T_{\text{single}} \approx \SI{8.1}{s}$. Consequently, for the long evolution times shown in the top panel, the individual ramsey contrasts  vanish and remain centered at zero. This confirms that the measured parity signal arises exclusively from the quantum correlations within the DFS. This result represents a coherence enhancement of nearly four orders of magnitude compared to the unprotected physical qubits, demonstrating the exceptional capability of the system to preserve quantum information over macroscopic timescales.

\begin{figure}[htbp]
\centering
\includegraphics[width=0.45\textwidth]{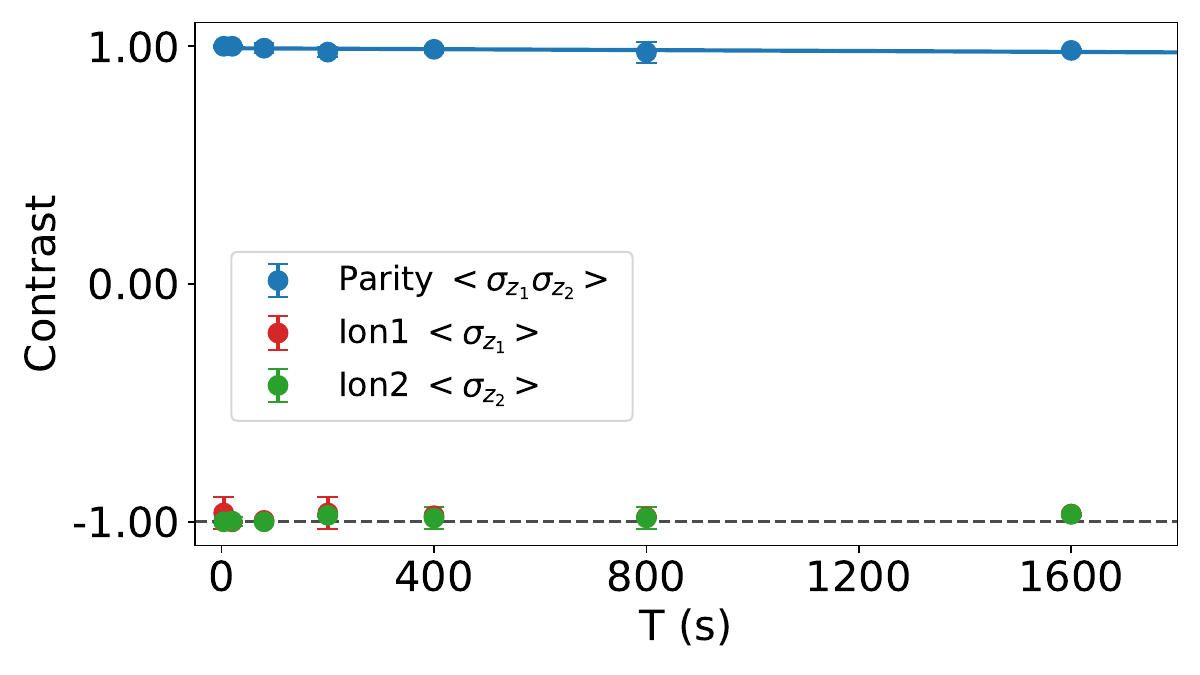}
\caption{\textbf{Lifetime measurement of the computational basis states.} The average parity contrast of the $\ket{0}\ket{0}$ state and  $\ket{1}\ket{1}$ state are monitored over time up to 1600s(blue circles). The data show no statistically significant decay, where exponential fits(solid blue line)  yield lower bounds on the lifetimes of $\rm T_1 = (1.01 \pm 0.71)\times 10^5$\,s. Contrast of individual qubits $\sigma_z$ are shown in red and green circles. }
\label{fig:t1_decay}
\end{figure}
To investigate depolarization time of $\rm T_1$, we monitor the lifetimes of the computational basis states $\ket{0}\ket{0}$ and $\ket{1}\ket{1}$ as an estimate. In Fig.~\ref{fig:t1_decay}, the plot displays the averaged population evolution for these states, where the single-ion observables are defined such that the target state corresponds to a value of $-1$. The data exhibit no statistically significant decay. An exponential fit to the parity yields an averaged of $\ket{0}\ket{0}$ and $\ket{1}\ket{1}$ lifetime of $\rm T_1 = (1.01 \pm 0.71)\times 10^5$\,s. 
This long lifetime confirms that the coherence time of the logical qubit is not limited by $\rm T_1$-time on this timescale.
 
\begin{figure}[htbp]
\centering
\begin{minipage}[t]{0.45\textwidth}
    \raggedright 
    \textbf{a}  
    \includegraphics[width=\textwidth]{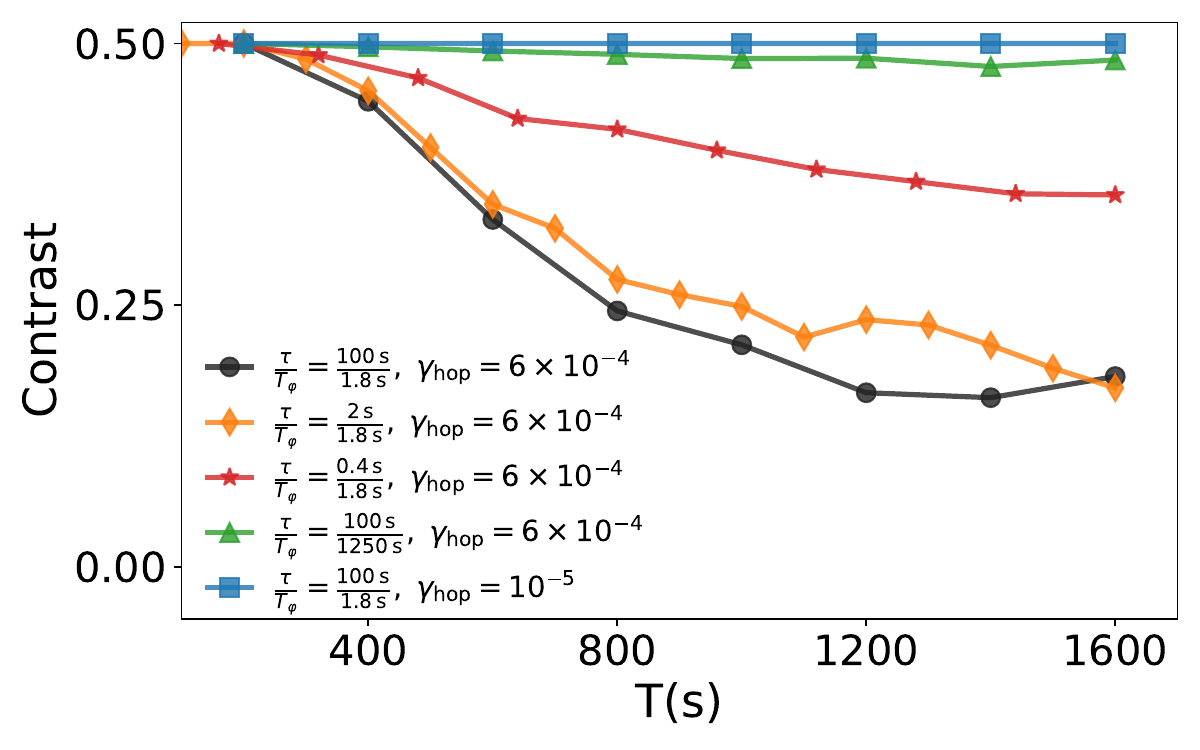}
\end{minipage}

\begin{minipage}[t]{0.45\textwidth}
    \raggedright
    \textbf{b} 
    \includegraphics[width=\textwidth]{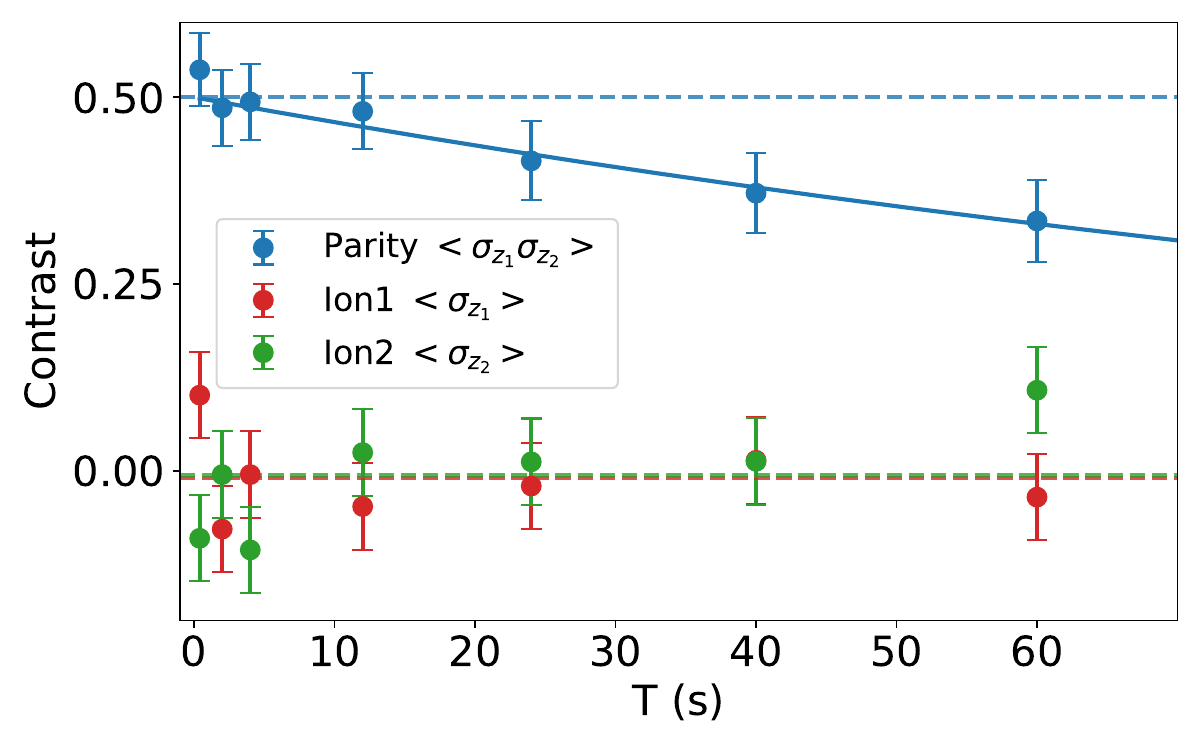}
\end{minipage}
\caption{\textbf{Investigation of coherence limiting mechanisms.} 
\textbf{(a)} Simulated coherence limits under varying ion hopping rates and magnetic field gradients, where the gradient intensity is parameterized by the DFS clock qubit phase evolution period $T_{\varphi}$. The black circles represents unoptimized conditions ($T_{\varphi}=\SI{1.8}{s},\gamma_{\rm hop}=6\times 10^{-4}$\,Hz). While direct hopping rate suppression ($10^{-5}$\,Hz, blue squares) yields optimal stability, suppressing the magnetic field gradient (increasing $T_{\varphi}$ to \SI{1250}{s}, green triangles) also substantially extends coherence, offering an effective alternative. Conversely, increasing spin-echo pulse density ($\tau=\SI{2}{s}$, orange; $\tau=\SI{0.4}{s}$, red) proves inefficient.
\textbf{(b)} Experimental DFS  Zeeman qubit decay. With an acquisition time far shorter than the hopping timescale, this measurement representing the coherence limit where hopping is negligible. The fitting result $\rm T_{\text{Zeeman}} = 145 \pm 14$\,s (blue line) allows projection of the magnetic-field-limited Clock coherence. Individual qubit means: $\braket{\sigma_{z1}} = -0.010\pm0.022$, $\braket{\sigma_{z2}} = -0.006\pm0.022$.}  
\label{fig:limitations}
\end{figure}

We note that leakage from laser and microwave fields makes a negligible contribution to the observed decoherence, owing to the high-extinction-ratio shuttering implemented in our setup~\cite{wang2021single} (see Methods). Instead, the primary limitation to the current coherence time of DFS logical qubit arises from stochastic exchange hopping of \Yb ions in the presence of a residual magnetic field gradient. This process flips the sign of the energy splitting, rendering the accumulated phase stochastic and irreversible by standard spin-echo techniques (see Methods). Numerical simulations, presented in Fig.~\ref{fig:limitations}a, illustrate this limitation. The  black circles represents the system prior to gradient compensation, characterized by a magnetic field gradient that results in a rapid phase evolution period of $T_{\varphi}=\SI{1.8}{s}$. In this regime, given our experiment spin echo interval $\tau=100~$s and the typical exchange hopping rate of $\gamma_{\text{hop}}\approx6 \times 10^{-4}$\,\si{Hz} is sufficient to randomize the  phase, severely limiting coherence to $10^3~$s level. We exclude exchange events between $^{171}\text{Yb}^+$ and $^{138}\text{Ba}^+$ ions from this analysis, as such configuration changes are detected and actively corrected by our real-time reordering protocol (see Methods).  

To address this, the simulations compare three mitigation strategies: (i) reducing the hopping rate (e.g., via cryogenic environments), (ii) reducing the interval $\tau$ of dynamical decoupling pulses, and (iii) suppressing the static magnetic field gradient. While reducing $\tau$ (orange and red curves) offers some protection, it is inefficient, and a large number of pulses leads to accumulation of coherent pulse errors. In contrast, the simulations identify that suppressing the magnetic field gradient is the suitable strategy for our room-temperature apparatus.

By employing the gradient compensation methodologies detailed in the Methods, we actively suppress the magnetic field inhomogeneity, effectively extending the phase evolution period $T_{\varphi}$ from \SI{1.8}{s} to \SI{900}\sim\SI{7000}{s}. This corresponds to a reduction in the magnetic field difference from $2.1\times 10^{-4}\,\text{G}$ to $4.2\times10^{-7} \sim 5.5\times10^{-8}\,\text{G}$. This reduction renders the system insensitive to our exchange hopping rate of $\gamma_{\text{hop}}= 6\times 10^{-4}$\,\si{Hz}. Specifically, it enables the use of pulse interval $\tau = \SI{100}{s}$ while still effectively suppressing phase errors from exchange hopping. Indeed, the simulation incorporating this suppressed gradient (Fig.~\ref{fig:limitations}a, green line) yields a predicted coherence time of $\rm T_{\text{sim}} = 3.95\times 10^4 $\,s,  consistent with our experimentally measured coherence time of $\rm T_{\text{DFS}} = (3.77 \pm 1.09)\times 10^{4}$\,s (Fig.~\ref{fig:clock_coherence}).

To evaluate the potential coherence time achievable in the absence of ion hopping, we characterize the magnetic field environment using a DFS Zeeman qubit~\cite{Kotler2013}, as shown in Fig.~\ref{fig:limitations}b. Since the data acquisition time for the Zeeman qubit is significantly shorter than the characteristic timescale of ion hopping ($\sim 10^3$\,s), the measurement effectively captures the dephasing caused by the residual magnetic field gradient in a static configuration. Experimentally, we apply a single reverse-type dynamical decoupling block and acquire data with 280 repetitions per point. An exponential fit to the decay yields a Zeeman coherence time of $\rm T_{\text{Zeeman}} = 145 \pm 14$\,s (blue line), where the uncertainty is derived from the fit statistics. The Zeeman qubit's energy splitting $\Delta E_Z$ is first-order sensitive to the gradient $\Delta B$ ($\Delta E_Z = \gamma_Z\Delta B$), whereas the Clock qubit's splitting $\Delta E_C$ depends on the much weaker differential second-order Zeeman shift ($\Delta E_C= \beta \Delta B^2\approx2\beta B\Delta B$). Given the meganetic field of $B \approx \SI{4.1}{G}$, the sensitivity ratio between the Zeeman and Clock qubits is calculated as $R = |\Delta E_Z|/|\Delta E_C| = \gamma_Z / (2\beta B) \approx 540$. Assuming the same underlying magnetic noise spectrum, we project a potential coherence time for the DFS Clock qubit of $\SI{78300}{s}$  (over 21 hours), which is calculated by $R \times \rm T_{\text{Zeeman}}$.

\section{Conclusion} 

In summary, our work establishes symmetry based protection through DFS as a potent form of quantum error correction. Within the framework of the Knill-Laflamme criteria, a DFS can be characterized as a highly degenerate quantum error correction code~\cite{PhysRevA.63.022307,PhysRevA.55.900}. This unique property ensures that the subspace serves as a simultaneous eigenspace for all Kraus operators within the correctable noise set, which in our system includes collective magnetic field fluctuations and common-mode local oscillator phase noise. Because these noise operators act as a scalar multiple of the identity on the logical manifold, the system satisfies the conditions for perfect error avoidance without the need for active syndrome measurements or unitary error reversal~\cite{PhysRevA.63.022307}. By encoding information into such a quiet corner of the Hilbert space, we achieve a robust architecture that remains completely immune to dominant environmental perturbations. This approach redefines the operational limits of room-temperature quantum memories and demonstrates that long coherence is accessible through the strategic suppression of addressable technical noise rather than being limited by fundamental atomic decoherence.

Looking ahead, the coherence times reported here are not fundamentally capped, with the remaining limits shown in Methods. By integrating cryogenic environments to suppress stochastic ion-exchange hopping~\cite{pagano_cryogenic_2018}, while utilizing magnetic shielding and stable field sources to ensure the spatial homogeneity and temporal stability of the magnetic gradient alongside individual beams for cooling ions, the decoherence rates could be pushed toward the natural limits of the atomic species. Under such improvements, coherence times extending to months or even longer may become achievable, finally approaching the million-year potential inherent in trapped-ion clock states. 

Such permanent quantum memory offers an unprecedented platform for high-precision metrology and the exploration of fundamental physical constants~\cite{RevModPhys.90.025008, PhysRevLett.113.210802}. To meet the demands of next-generation scalable architectures~\cite{monz2009realization}, DFS encoding provides a robust solution, particularly for QCCD systems where qubits must idle in memory zones for extended periods~\cite{ransford2025helios98qubittrappedionquantum}. By anchoring protection to the relative energy of ion pairs, this method shields logical information from transport-induced noise during shuttling, effectively bridging the gap between extreme coherence and system scalability. Such a foundation is vital for global quantum networks and distributed computing~\cite{wehner_quantum_2018}, as it enables the preservation of entanglement over timescales exceeding classical communication latencies and satellite orbital periods, ultimately paving the way for a truly persistent quantum information infrastructure.



\section{Data availability}
The data supporting this work are available from the authors upon request.

\bibliography{references}

\section{Methods}


\subsection{Mitigation of ion-hopping-induced decoherence via gradient suppression}

\begin{figure}[htbp]
    \centering
    \begin{minipage}[t]{0.48\textwidth}
        \raggedright
        \textbf{(a)}
        \includegraphics[width=\textwidth]{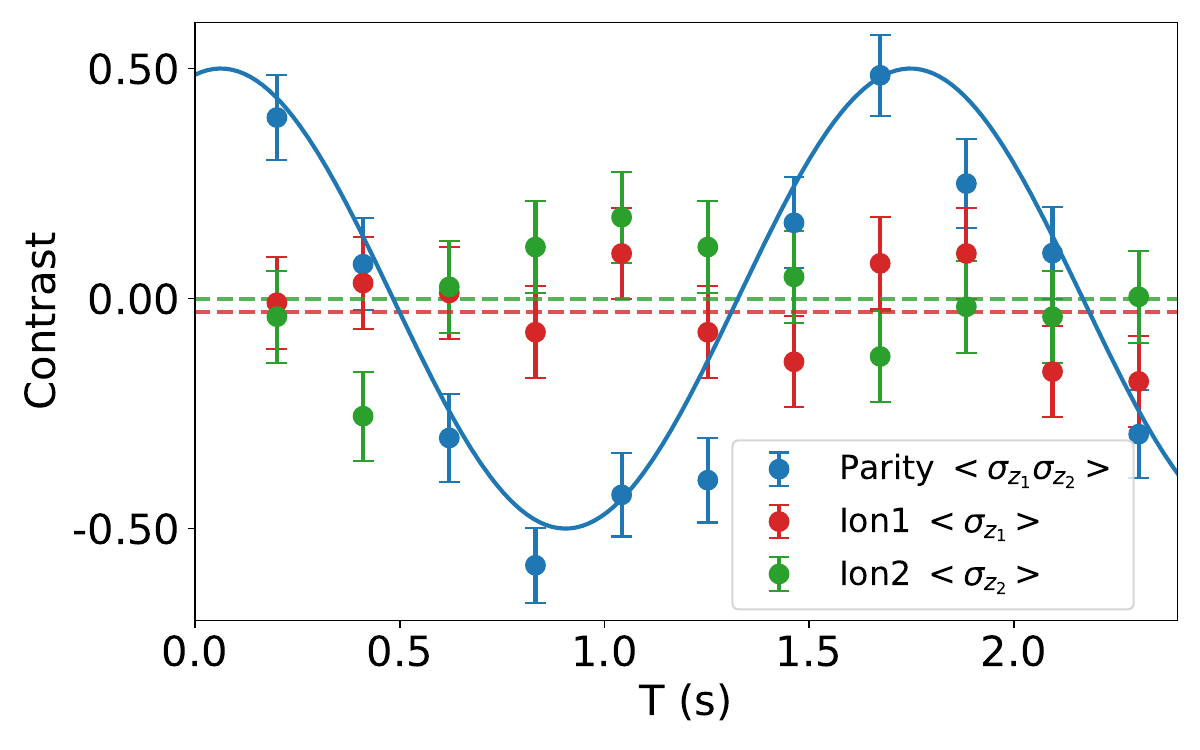}
    \end{minipage}
    \hfill
    \begin{minipage}[t]{0.48\textwidth}
        \raggedright
        \textbf{(b)}
        \includegraphics[width=\textwidth]{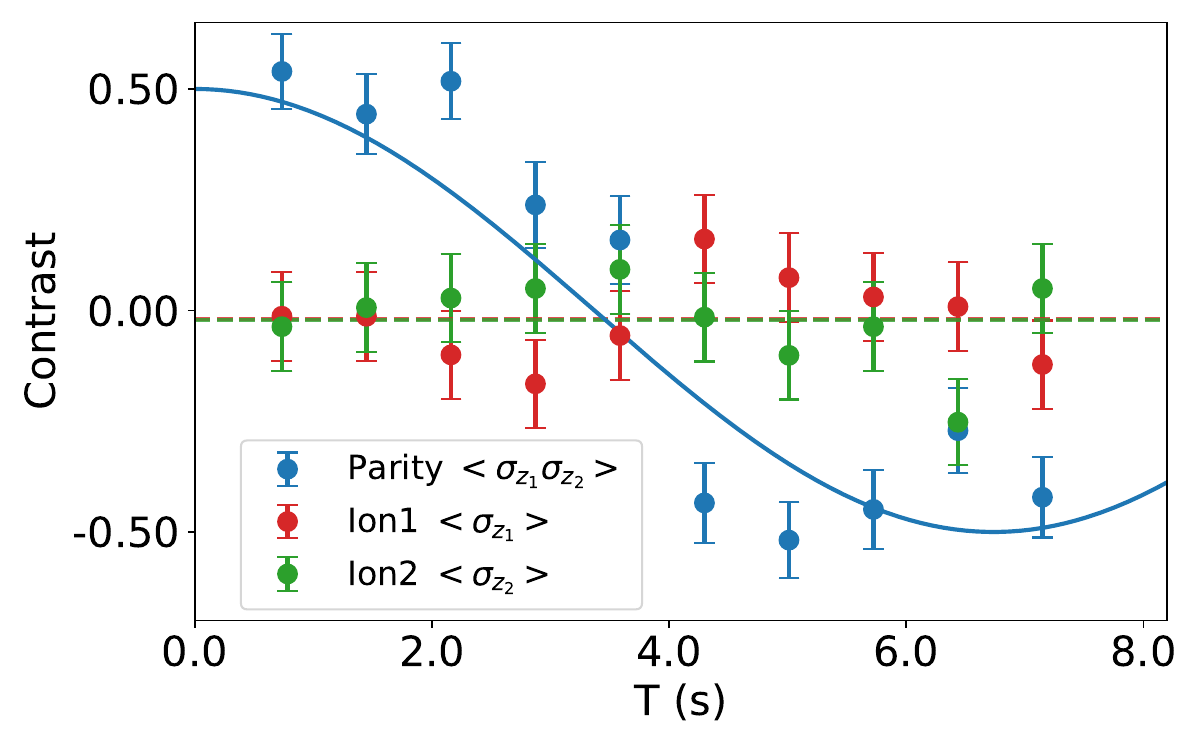}
    \end{minipage}
    \caption{Characterization of the suppressed magnetic field gradient via Zeeman DFS oscillations. While an ideal zero-gradient environment would yield no oscillation, the long periods observed here demonstrate a highly homogeneous field achieved via active compensation. 
\textbf{(a)} A representative lower bound for the oscillation period ($T_{\varphi}^{Zeeman} = 1.68 \pm 0.05$\,s) observed under typical compensated conditions. 
\textbf{(b)} An attained upper bound showing a period of $T_{\varphi}^{Zeeman} = 13 \pm 1$\,s. 
Compared to the uncompensated period of $\approx 3$\,ms, these results represent an improvement of over three orders of magnitude.}
    \label{fig:zeeman_oscillation}
\end{figure}
A dominant decoherence mechanism for our logical qubit is the stochastic exchange hopping of ions in the presence of a residual magnetic field gradient. The field difference $\Delta B$ proportionally induces an energy splitting $\Delta E $ between the DFS basis states, causing the logical qubit to accumulate a deterministic phase $\varphi(t) = \Delta E \cdot t / \hbar$ during free evolution. In a standard spin-echo sequence with interval $\tau$, a $\pi$-pulse reverses this accumulation. However, if the ions randomly exchange positions at time $t_{hop}$, the sign of the splitting flips ($\Delta E \to -\Delta E$). The net accumulated phase error after the interval $\tau$ becomes:
\begin{equation}
    \varphi_{accu} = \int_{n\tau}^{t_{hop}} \frac{\Delta E}{\hbar} dt + \int_{t_{hop}}^{(n+1)\tau} \frac{-\Delta E}{\hbar} dt .
\end{equation}
Since $t_{hop}$ is a random variable, $\varphi_{accu}$ is stochastic and cannot be perfectly cancelled by the spin-echo pulse, leading to dephasing.

The magnitude of this error is governed by the ratio $\tau/T_{\varphi}$, where $T_{\varphi} \propto \hbar/\Delta E$ is the DFS free-evolution period. To mitigate this, one must minimize $\tau/T_{\varphi}$. Three strategies exist: (i) reducing the hopping rate (e.g., using a cryogenic trap) to make hopping events rare; (ii) reducing $\tau$ by increasing the number of spin-echo pulses; or (iii) increasing $T_{\varphi}$ by suppressing the gradient $\Delta B$. Our numerical simulations (Fig.~\ref{fig:limitations}a) indicate that while cryogenics is effective, it is not feasible for our room-temperature setup. Merely increasing the pulse number to reduce $\tau$ is inefficient, as the accumulation of pulse errors eventually degrades coherence. Therefore, the optimal strategy is to suppress the magnetic field gradient, thereby maximizing $T_{\varphi}$. This drastically slows the phase evolution rate, rendering the phase error accumulated during any finite hopping interval negligible.

To implement this strategy, we generate a uniform magnetic field along the x-direction using a large ring-shaped permanent magnet. The average field is calibrated to $B \approx \SI{4.1}{G}$. Residual gradients in the z and y directions are actively nulled using two small compensation magnets. The effectiveness of this compensation is characterized by the DFS Zeeman qubit oscillation period, which serves as a sensitive probe of the gradient. Before compensation, the period was a mere \SI{3.0}{ms} ($\Delta B \approx \SI{2.4e-4}{G}$). After compensation, the period extended to typically $T_{\varphi}^{Zeeman} \approx \SI{1.68}{s} \sim  \SI{13}{s}$ (Fig.~\ref{fig:zeeman_oscillation}), representing an improvement of nearly three orders of magnitude to the $4.2\times10^{-7} \sim 5.5\times10^{-8}$ G level. Although the residual gradient exhibits slow temporal drifts, the oscillation period consistently remains above 1.68 second. This robust suppression justifies the conservative estimate used in our simulations, where a Zeeman period on the order of seconds implies a DFS clock qubit period exceeding 900 seconds (scaling by the sensitivity ratio of $\approx 540$).


\subsection{Suppression of microwave and 369 laser leakage}

\begin{figure*}[htbp]
\centering
\includegraphics[width=0.8\textwidth]{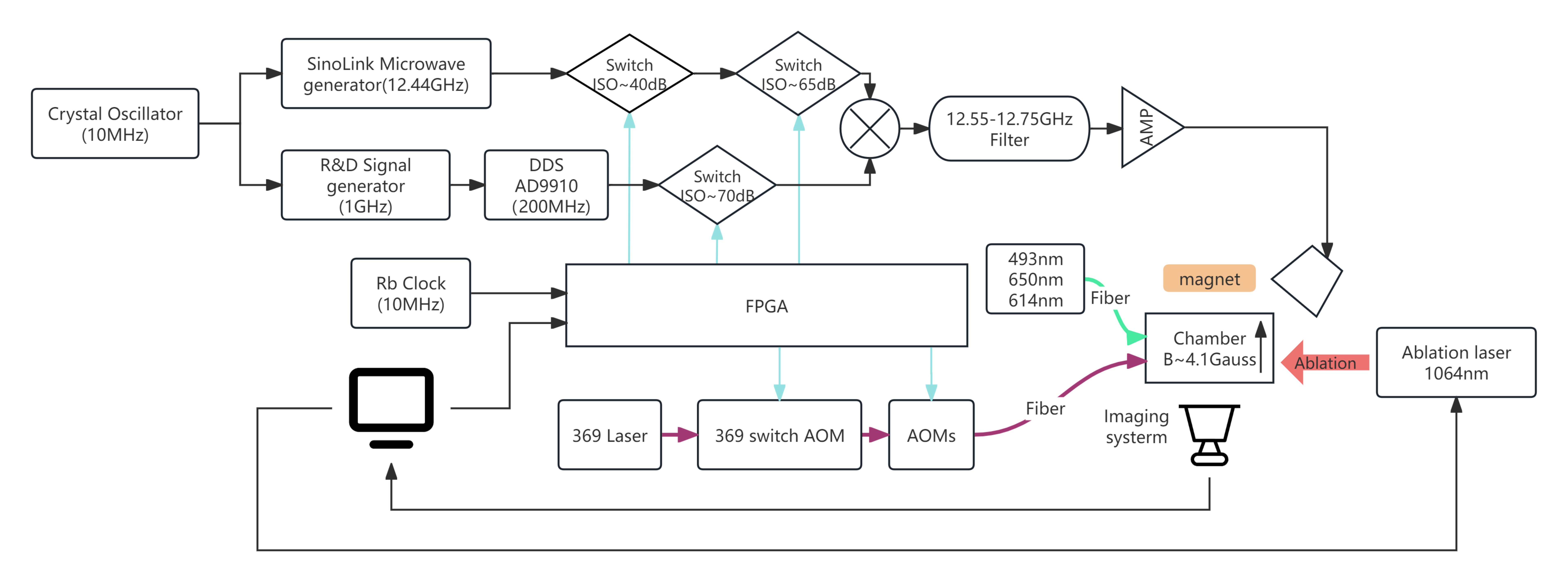} 
\caption{Schematic of the experimental control system. A host computer and an FPGA orchestrate all timing sequences. The \SI{12.64}{GHz} microwave signal for qubit manipulation is synthesized by mixing a \SI{12.44}{GHz} microwave oscillator with a \SI{200}{MHz} signal from a DDS. All sources are phase-locked to a \SI{10}{MHz} crystal oscillator for phase stability. A multi-stage switching network provides high isolation to suppress microwave leakage. Laser beams are controlled by acousto-optic modulators (AOMs) for rapid switching and are delivered to the ion trap via optical fibers.}
\label{fig:setup} 
\end{figure*}

In our experimental setup, considerable effort is devoted to minimizing optical and microwave leakage. To suppress stray 369 laser, a high-extinction-ratio shutter system is implemented, consisting of two cascaded acousto-optic modulators (AOMs) along with two spatial filter apertures. The beam is then coupled into a single-mode optical fiber, which serves to refine the spatial mode profile and suppress off-axis scattered laser.
\par
Microwave leakage presents a comparable challenge in our experimental setup. The \SI{12.64}{GHz} microwave field used to drive Rabi oscillations is generated by mixing a \SI{200}{MHz} signal from a direct digital synthesizer (DDS) with a \SI{12.44}{GHz} microwave oscillator, as shown in Fig.~\ref{fig:setup}. To ensure high phase stability and spectral purity, all microwave frequency sources are phase-locked to a common \SI{10}{MHz} crystal reference oscillator; the DDS is specifically clocked by a \SI{1}{GHz} signal derived from this same reference to get lowest phase noise. A narrow-band pass filter selects the desired \SI{12.64}{GHz} component. To suppress driving caused by microwave leakage, a multi-stage switching network is employed. This includes a fast switch offering \SI{70}{dB} isolation installed at the DDS output. Simultaneously, the \SI{12.44}{GHz} signal path is gated using two cascaded switches, providing isolations of \SI{40}{dB} and \SI{65}{dB}, respectively. Together, these measures effectively suppress microwave leakage to a negligible level throughout the experimental sequence. No significant microwave or laser leakage was observed over the $1,600~\text{s}$ measurement period, as evidenced by the lack of discernible decay in the $\ket{1}$ state population for each qubit when no operations were applied.

\subsection{Benchmark of Qubit Control}
\begin{figure}[htbp]
    \centering
    \subfigure[]{\includegraphics[width=0.45\textwidth]{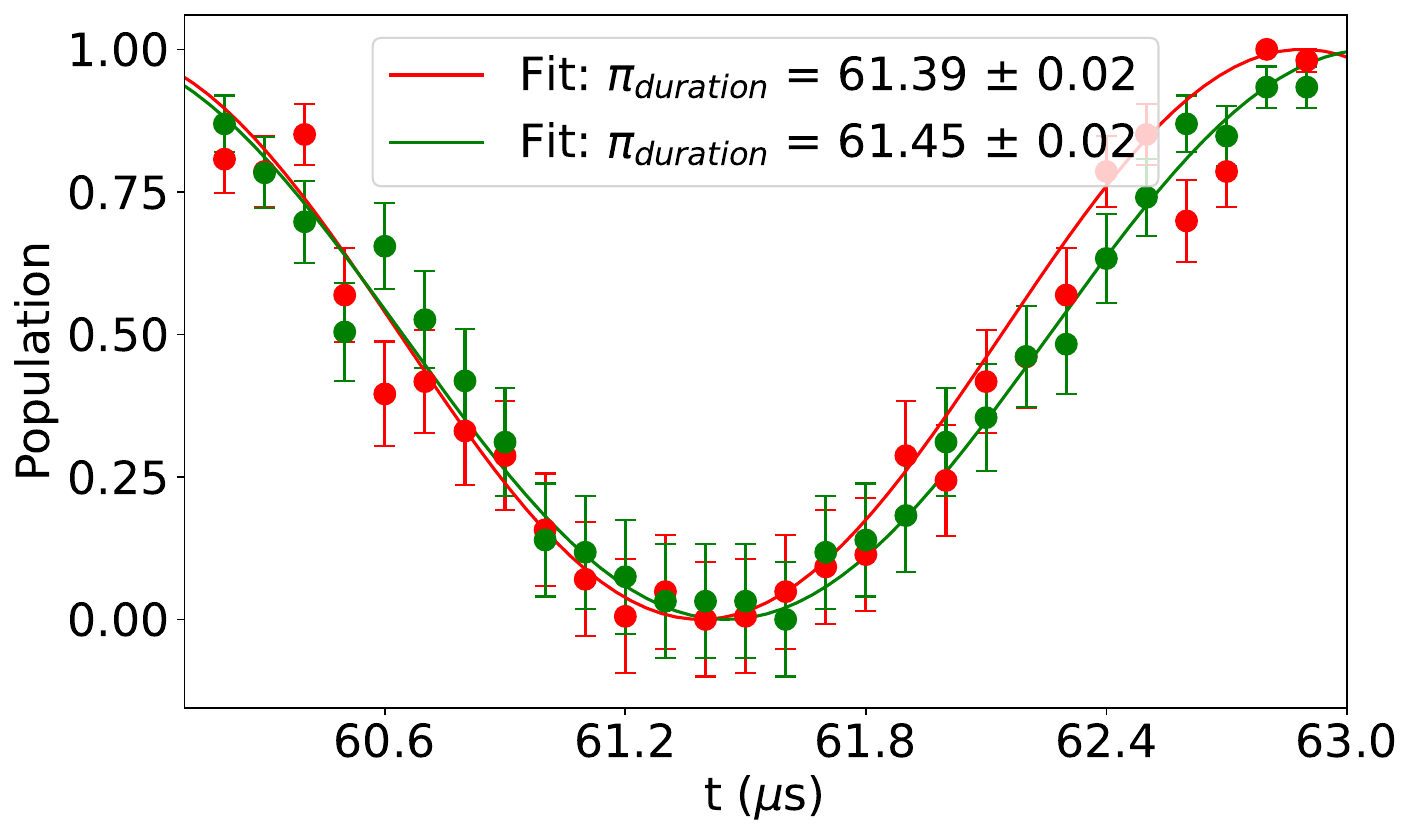}\label{fig:pi_calibration}}
    \subfigure[]{\includegraphics[width=0.45\textwidth]{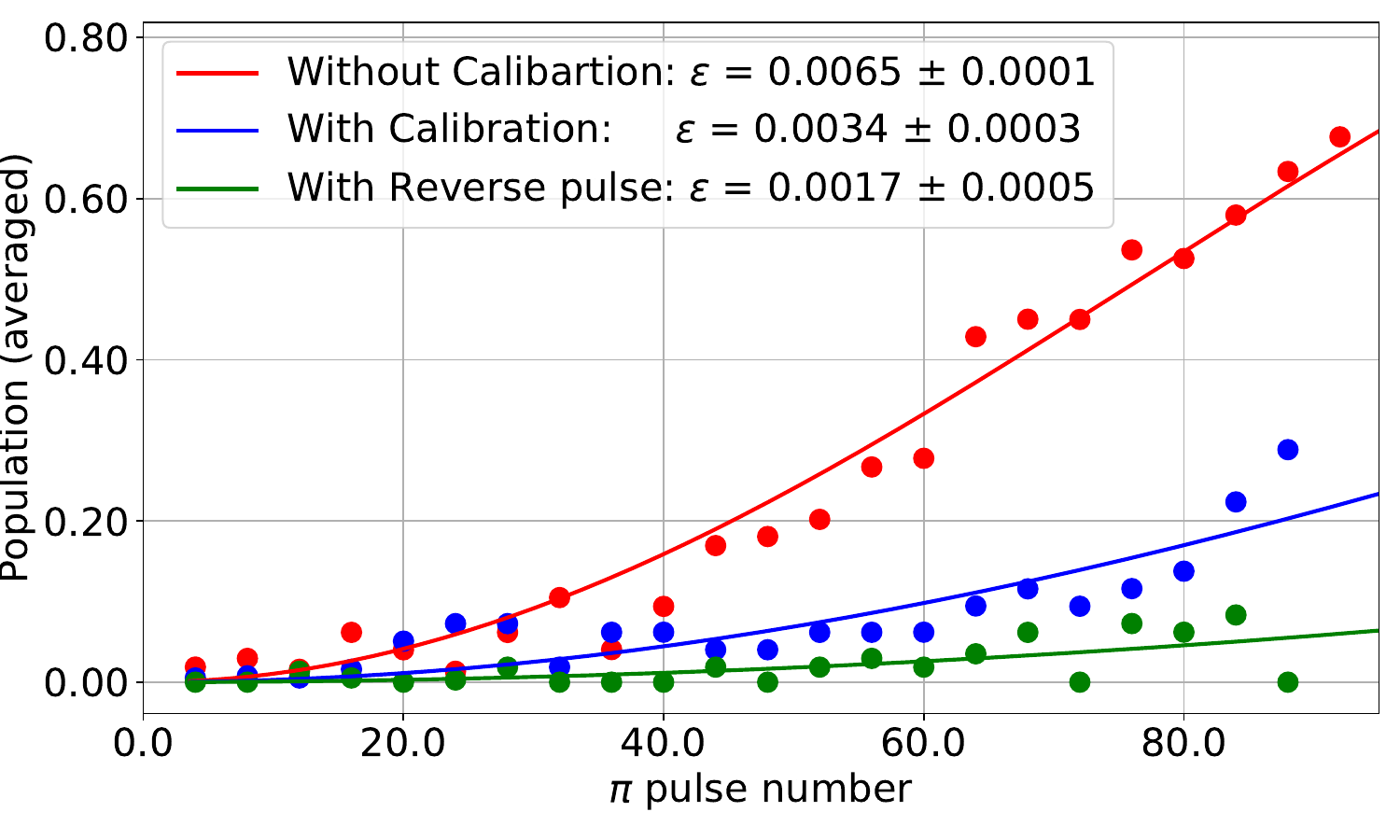}\label{fig:benchmark_error}}
    \caption{Calibration and benchmarking of the microwave $\pi$-pulse.
    \textbf{(a)} High-precision calibration of the $\pi$-pulse duration for each ion using an error-amplification sequence. The minima of the curves indicate the optimal gate times.
    \textbf{(b)} Benchmarking of the accumulated population error as a function of the number of $\pi$ pulses. Comparison of uncalibrated (red), calibrated (blue), and reverse-type (green) sequences, where $\epsilon$ denotes the relative pulse error. The suppressed $\epsilon$ in the green curve demonstrates robust protection against systematic over-rotations. Data represent the average of two ions.}
    \label{fig:pi_pulse_benchmark} 
\end{figure}
Precise calibration of the microwave $\pi$-pulse duration is critical for high-fidelity operations. To achieve this, an error amplification sequence consisting of 40 consecutive $\pi$-pulses is applied to each qubit initialized in the $\ket{0}$ state. As shown in Fig.~\ref{fig:pi_calibration}, this method allows the optimal $\pi$-time to be determined with a precision of \SI{0.02}{\micro\second}. By carefully aligning the microwave horn to homogenize the field across the ion crystal, the calibrated durations for the two ions are made nearly identical: $\pi_{duration, 1} = 61.39 \pm 0.02$\,\textmu s and $\pi_{duration, 2} = 61.45 \pm 0.02$\,\textmu s.

Furthermore, to enhance the robustness of our main experimental sequence against residual calibration errors and power fluctuations, the spin-echo blocks are constructed from pairs of $\pi$-pulses with opposite phases,  which we name as reverse style spin-echo. The effectiveness of this composite pulse design is benchmarked in Fig.~\ref{fig:benchmark_error}. The data demonstrate that a simple sequence without compensation (red) accumulates errors quadratically with pulse length offsets. While precise calibration significantly reduces this error (blue), the composite pulse sequence (green) offers the most robust performance, suppressing the error by an additional factor of two and achieving a minimal average relative pulse error of $\epsilon = 0.0017 \pm 0.0005$.

\subsection{Stabilization and reordering of ion configuration}

\begin{figure}[htbp]
\centering
\includegraphics[width=0.5\textwidth]{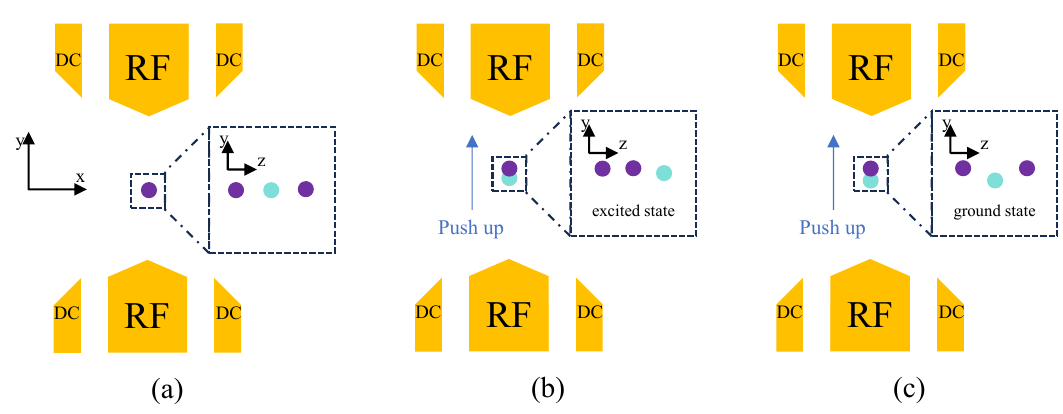}
\caption{Ion configuration reordering protocol. 
\textbf{(a)} The target Yb--Ba--Yb configuration. 
\textbf{(b)} An excited crystal configuration (e.g., Yb--Yb--Ba) in the presence of micromotion. A controlled perturbation induces relaxation from this micromotion-dressed excited state to the ground state. 
\textbf{(c)} The ground-state crystal configuration. This state can be adiabatically returned to the target centered position (a) by slowly ramping down the DC offset voltage.}
\label{fig:reorder-ion}
\end{figure}

Maintaining the target Yb--Ba--Yb crystal configuration is essential for stable quantum memory operation, as background gas collisions can intermittently disrupt the ion arrangement. In our chip-based ion trap, we load two $^{171}\text{Yb}^+$ ions and one $^{138}\text{Ba}^+$ ion via laser ablation to form the target symmetric configuration (Fig.~\ref{fig:reorder-ion}a). Despite the high-vacuum environment, rare but inevitable collisions with residual molecules can swap the ion positions, necessitating a robust reordering protocol~\cite{PhysRevLett.134.023201}.

The stability of specific ion arrangements is dictated by mass-dependent radial trap frequencies in the presence of micromotion. Because the radial confinement scales as $\omega_{x,y} \propto 1/M$, ions of different masses experience distinct effective potentials. Consequently, the Yb--Ba--Yb arrangement represents the ground-state configuration (Fig.~\ref{fig:reorder-ion}b), while asymmetric arrangements such as Yb--Yb--Ba correspond to higher-energy excited states \cite{James1998}. By introducing a controlled perturbation, the system can be prompted to relax from these metastable excited states back to the ground-state configuration.

We implemented an automated detection and correction system to ensure the ion crystal remains in the desired configuration throughout long-term measurements. The position of the $^{138}\text{Ba}^+$ ion is monitored in real time using a dedicated EMCCD camera, where any significant displacement triggers an automatic recovery sequence. Once a configuration error is detected, the ion crystal is adiabatically shifted off-center. The RF power is then abruptly switching to a lower power level and back. This perturbation facilitates the relaxation to the ground state, after which the crystal is adiabatically returned to the trap center.

To ensure data integrity, we utilize simultaneous dual-species imaging to filter out measurements affected by configuration errors or ion loss. Two QCMOS cameras independently monitor the $^{171}\text{Yb}^+$ and $^{138}\text{Ba}^+$ ions; any data point where the $^{138}\text{Ba}^+$ photon count falls below a pre-calibrated threshold is automatically discarded. This post-selection ensures that all reported coherence data are derived exclusively from the correctly ordered Yb--Ba--Yb crystal.

\subsection{Limitation of coherence time}

\begin{figure}[htbp]
\centering
\includegraphics[width=0.45\textwidth]{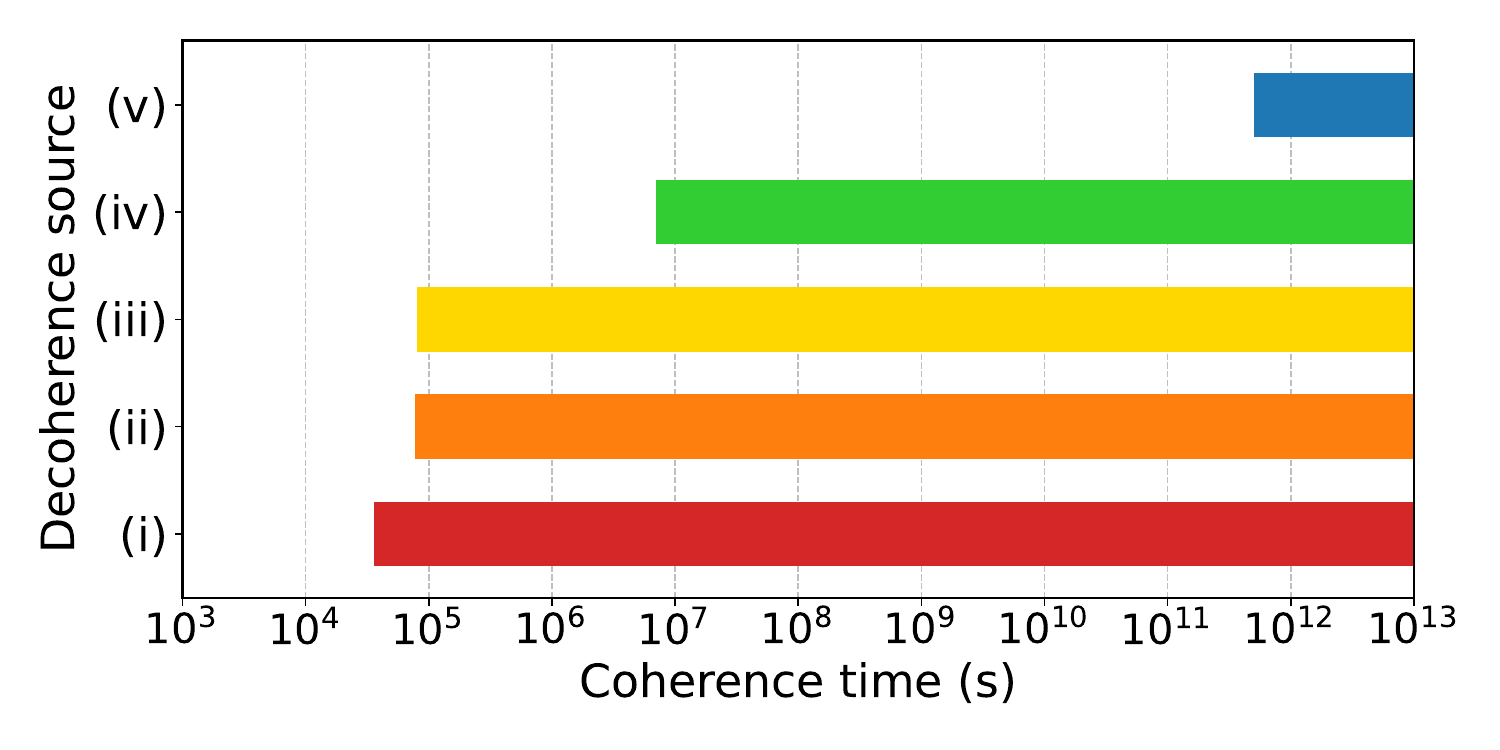}
 \caption{\textbf{Expected limitations of coherence time with DFS clock qubit.} The estimated coherence time limits imposed by various noise sources are presented, ranked from most to least detrimental: (i) Stochastic ion exchange hopping, limited by the residual field gradient ($\sim 3.8 \times 10^4$\,s); (ii) Residual magnetic field fluctuations ($\delta B$), projected from Zeeman measurements ($\sim 7.8 \times 10^4$\,s); (iii) Off-resonant photon scattering from the sympathetic cooling lasers ($\sim 8 \times 10^4$\,s); (iv) Pulse Imperfections, suppressed benchmark and reverse type spin-echo ($\sim 7 \times 10^6$\,s); and (v) The fundamental $T_1$ lifetime of the ground state ($\sim 5 \times 10^{11}$\,s). The left edge of each bar represents the estimated coherence limit, illustrating that ion hopping remains the primary bottleneck for our current apparatus.}
\label{fig:decohrence_analy}
\end{figure}

To understand the ultimate limits of our quantum memory, we systematically analyze the hierarchy of decoherence mechanisms, as summarized in Fig.~\ref{fig:decohrence_analy}. The estimated coherence limits imposed by these sources are discussed below in order of their impact.

\textbf{Stochastic Ion Exchange Hopping:} The primary limitation to the current coherence time is identified as stochastic ion hopping. Our experimental measurements yield a coherence time of $\sim 3.8 \times 10^4$\,s. This result aligns well with our numerical simulations (see Fig.~\ref{fig:limitations}), which indicate that the residual magnetic field gradient, combined with the experimentally measured hopping rate ($\sim 10^{-3}$ Hz), limits the coherence to approximately this same level. This agreement confirms that stochastic ion hopping in the presence of a residual gradient is the dominant error source limiting the system's performance. Implementing a cryogenic trap would suppress the ion hopping rate to $10^{-5}$Hz~\cite{pagano_cryogenic_2018}, potentially increasing the coherence time by two orders of magnitude.  

\textbf{Residual Magnetic Field Noise:} The second most significant factor is the residual magnetic field noise acting on the DFS clock qubit. While the DFS suppresses common-mode noise, it remains sensitive to fluctuations in the magnetic field gradient. Based on the measured coherence time of the sensitive DFS Zeeman qubit($\rm T_{Zeeman} \approx 145$\,s) and the sensitivity ratio $R \approx 540$, the projected limit for the DFS clock qubit due to magnetic field noise alone is $\rm T_{Clock} \approx 7.8 \times 10^4$\,s. This indicates that magnetic field stability is close to becoming a limiting factor alongside ion hopping. Installing a magnetic shield typically suppresses magnetic field noise by 20 dB~\cite{wang2021single}.

\textbf{Off-Resonant Photon Scattering:} A fundamental limit arises from the off-resonant scattering of the $^{138}\text{Ba}^+$ cooling lasers by the $^{171}\text{Yb}^+$ qubits. We estimate this rate using a Raman scattering model:
\[
    \Gamma_{\text{scat}} = \frac{g^2\Gamma}{6} \left( \frac{1}{\Delta_{D1}^2} + \frac{2}{(\Delta_{D1} + \Delta_{fs})^2} \right),
\]
where $\Gamma \approx 2\pi \times \SI{20}{MHz}$ is the natural linewidth and $\Delta_{fs} \approx 2\pi \times \SI{100}{THz}$ is the fine-structure splitting. The coupling strength $g = \frac{\Gamma}{2}\sqrt{I / (2I_{\text{sat}})}$ depends on the laser intensity $I$ at the ion's position. For our setup, the beams are incident at $45^\circ$, reducing the effective intensity by a factor of $1/\sqrt{2}$. Additionally, since the cooling beams are aligned to the central $^{138}\text{Ba}^+$ ion, the $^{171}\text{Yb}^+$ ions are displaced from the beam center by the experimentally measured ion separation of $r \approx \SI{6}{\micro\meter}$.

For the \SI{493}{nm} laser ($P = \SI{220}{\micro\watt}$, waist $w = \SI{38}{\micro\meter}$), the calculated scattering rate is $\Gamma_{493} \approx 6.06 \times 10^{-6}$\,Hz. For the \SI{650}{nm} repumper ($P = \SI{90}{\micro\watt}$, waist $w = \SI{55}{\micro\meter}$), the rate is $\Gamma_{650} \approx 4.76 \times 10^{-7}$\,Hz. Since the logical qubit involves two $^{171}\text{Yb}^+$ ions, the total decoherence rate is doubled: $\Gamma_{\text{total}} \approx 2 \times (\Gamma_{493} + \Gamma_{650}) \approx 1.3 \times 10^{-5}$\,Hz. Accounting for uncertainties in beam alignment, this imposes a coherence limit in the range of $8 \times 10^4$\,s to $1.1 \times 10^5$\,s. While not the current bottleneck, this limit could be further extended by reducing the cooling laser power or implementing individual addressing for the Ba ion.

\textbf{Pulse Imperfections:} While systematic pulse errors are effectively compensated by our reverse-phase pulse sequence, residual stochastic fluctuations in the relative pulse error $\epsilon$ impose a technical upper bound on coherence. Each decoupling pulse introduces a random rotation angle deviation $\delta\theta \approx \pi\epsilon$, which accumulates as a random walk over a sequence of $N$ pulses. The resulting total angular variance, $\langle \Theta^2 \rangle = N(\pi\epsilon)^2$, leads to an exponential decay of the coherence contrast $C \propto \exp(-\frac{1}{2} \langle \Theta^2 \rangle)$. Given our measured stability ($\epsilon \approx 1.7 \times 10^{-3}$) and a pulse interval of $\tau = \SI{100}{s}$, we estimate the timescale for this accumulated angular error to reach the $1/e$ threshold to be $T \approx 7.0 \times 10^6$\,s. The impact of $\epsilon$ could be further suppressed by employing more sophisticated decoupling sequences, such as the Knill dynamical decoupling (KDD) protocol, which provides higher-order robustness against systematic pulse imperfections. 

\textbf{Microwave Leakage:} Microwave leakage can drive unintentional qubit rotations, but is heavily suppressed in our setup by a multi-stage switching network providing approximately \SI{175}{dB} of isolation. At this level of attenuation, the rotation angle accumulated due to the residual leakage field over a \SI{100}{s} interval is approximately $1.8 \times 10^{-2}$ radians (calculated relative to the \SI{61}{\micro s} $\pi$-pulse duration). This accumulated rotation is actively cancelled by inverting the phase of the microwave pulses in our dynamical decoupling sequence. However, the precision of this cancellation is limited by the relative pulse error ($\epsilon \approx 1.7\times10^{-3}$), leading to a residual angular error that accumulates stochastically. We estimate the resulting coherence limit to be on the order of $8.2 \times 10^{11}$\,s, rendering microwave leakage completely negligible for our measurements. 

\textbf{Spontaneous Emission ($T_1$):} Finally, the fundamental spontaneous emission lifetime of the $^{171}\text{Yb}^+$ ground state hyperfine levels sets an ultimate upper bound. This $T_1$ time is theoretically estimated to be extremely long ($> 10^{11}$\,s)~\cite{wild1952radio,4065250,pritchard201221}, corresponding to a limit of $\sim 5 \times 10^{11}$\,s for our system, which is effectively infinite for quantum information purposes.

\section*{Acknowledgements}
This work was supported by the Quantum Science and Technology-National Science and Technology Major Project No. 2021ZD0301602 and the National Natural Science Foundation of China under Grants No. 92065205, 92576204, No. 92565306, No. 11974200, No. 62335013, and No. 12304551
KK acknowledges the support by the IBS Grant No. IBS-R041-D1.

\section{Ethics declarations}
No competing interest.

\section*{Author Contributions}
K.K. conceived the experiment and the DFS encoding scheme. X.J.L., J.H.P. and J.L.C. built the experimental apparatus and the ion-trap system. J.H.P. and X.J.L. performed the experiment measurements and analyzed the data. X.Z. and J.H.Z provided support for the FPGA infrastructure. L.F.O. and E.F.G. provided the ion-trap chips. P.F.W., H.C.T., and M.L.Z. participated in scientific discussions and provided technical assistance. J.H.P., X.J.L., and K.K. wrote the manuscript with input from all authors.
\end{document}